\documentclass[a4paper,12pt]{article}
\usepackage{graphicx}
\usepackage{amsmath}

\begin{document} 

\title{MAPS OF ZEROES OF THE GRAND CANONICAL PARTITION FUNCTION IN A STATISTICAL MODEL OF HIGH ENERGY COLLISIONS}
\author{M. Brambilla (*), A. Giovannini (**), R. Ugoccioni (**)\\
 \textit{\small(*) Dipartimento di Fisica Nucleare e Teorica and INFN - Sez. di Pavia}\\
 \textit{\small Via A. Bassi 6, 27100 Pavia, Italy} \\
 \textit{\small(**) Dipartimento di Fisica Teorica and INFN - Sez. di Torino}\\
 \textit{\small Via P. Giuria 1, 10125 Torino, Italy}}
\maketitle

\begin{abstract}
Theorems on zeroes of the truncated generating function in the
complex plane are reviewed. When examined in the framework of a statistical
model of high energy collisions based on the negative binomial
(Pascal) multiplicity distribution, these results lead to maps of zeroes of the
grand canonical partition function which allow to interpret in a novel
way different classes of events in pp collisions at LHC c.m.\
energies.
\end{abstract}

\section{Introduction}    

Following previous work \cite{thermo-clan}, the multiplicity
distribution (MD) generating function in
the dummy variable $z$, 
\begin{equation}
	G(z)=\sum_{n=0}^{\infty}P_{n}z^n     \label{eq:1}
\end{equation}
has been recognised to be related to the grand canonical partition
function of a system of $n$ particles in statistical mechanics when
$z$ 
is identified with the fugacity variable; $P_{n}$ is here the
probability of detecting $n$ charged particles in full phase space
($\sum_{n=0}^{\infty}P_{n}=1$) and it is linked to the canonical
partition function.  

It has to be stressed that in a realistic experiment $n$ will never
become infinity. Let us call $M$ the maximum finite 
value of $n$ which can be detected in an experiment. The generating
function $G(z)$ in this case is reduced to a polynomial of degree $M$
in $z$: 
\begin{equation}
	G_{M}(z)=\sum_{n=0}^{M}P_{n}z^n.     \label{eq:2}
\end{equation}

Let us now consider the algebraic equation
\begin{equation}
	G_{M}(z)=0     \label{eq:3}
\end{equation}
in the complex $z$-plane.
In the first place, notice that none of the roots $z_i$ $(i=1, \dots,
M)$ of equation (\ref{eq:3}) 
can be real and positive, since the $P_{n}$ in (\ref{eq:2}) are positive for
any $n$. In the second place, recall that the
$M$-truncated generating function (as any polynomial) 
can be factorised in terms of its roots as follows:
\begin{equation}
	G_{M}(z)=P_{M}\prod_{i=1}^{M}(z-z_{i}).     \label{eq:4}
\end{equation}

By applying these considerations to phase transitions of a lattice gas
(and of the Ising model), C.N.\ Yang and T.D.\ Lee \cite{Lee-Yang} have
found that the roots $z_i$ 
lie on a circle centered at the origin of the complex $z$-plane. The
circle is open in a small sector bisected 
by the positive real axis; no zero lies in that sector. When $M$ is
odd, at least one root is real: it will of course be on the negative 
side. $G_{M}(z)$ turns out to be an analytic function along the
positive real axis for any $M$. A phase transition in the sense 
of a non analytic behaviour of the thermodynamical functions (like
pressure for instance) can only occur at points of the positive 
real $z$ axis which are ``in the thermodynamical limit'' accumulation
points of the zeroes of the algebraic equation (\ref{eq:3}).  
For $M \rightarrow \infty$, one should expect in  general that the
zeroes become closer and closer to the real positive axis and  
that the distance between the zeroes vanishes.

The first application of these ideas to particle production 
goes back to the 70's \cite{Biebl}, but the field flourished again
(results were obtained in the context a discrete approximation to QCD
cascades \cite{Discrete})
in 1995, when E.\ De Wolf  \cite{De Wolf}  studied,
using the JETSET Monte Carlo generator,
the zeroes of the generating function of the $n$-charged particle
multiplicity distribution in $e^{+}e^{-}$ annihilation in the
$y$-rapidity interval $|y| < 0.5$ at 1000 GeV c.m.\ energy.
The zeroes were found to lie on a circle of approximately unit radius
centered at the origin of the complex $z$-plane. 
Intriguing questions were also raised on the development 
of this approach
in different models
and in different classes of events.

After recalling and extending the main theorems on the subject,
the present paper explores the properties of the
distribution of zeroes in high energy collisions
in the framework of the weighted superposition mechanism of different
classes of events: each class is described by a Negative Binomial
(Pascal) MD, whose characteristic parameters are $\bar n$, the average
charged particle multiplicity, and $k$, linked to the dispersion $D$
by $k={\bar n}^2/(D^2-\bar n)$.

\section{Maps of zeroes in the complex plane and the class of power series distributions.}

In 1997, T.C.\ Brooks, K.L.\ Kowalsky and C.C.\ Taylor
\cite{Brooks} 
applied the Enestr\"{o}m-Kakeya theorem \cite{Marden}
to the scaled $M$-truncated generating function 
\begin{equation}
	G_{M}(\lambda z)=\sum_{n=0}^{M}c_{n}z^n     \label{eq:5}
\end{equation}
with coefficients given by $c_{n}=\lambda^{n}P_{n}$ and the parameter
$\lambda$ defined as the minimum of the ratio of two consecutive $c_{n}$
 coefficients:
\begin{equation}
	\lambda = r[n_{\min}]=\min[P_{n}/P_{n+1}]=P_{n_{\min}}/P_{n_{\min}+1}.     \label{eq:6}
\end{equation}

The coefficients $c_{n}$, for $n=0,1, \dots, M$, are real and positive
and satisfy the condition
$$
c_{0}\geq c_{1}\geq \dots \geq c_{M}
$$ 
required by the Enestr\"{o}m-Kakeya (EK) theorem in order to ensure
that $G_{M}(\lambda z)$ differs from zero in the complex $z$-plane for
$|z| < 1$.  
Equation (\ref{eq:5}) can be rewritten in terms of the variable $w=1/z$ as 
\begin{equation}
	G_{M}(\lambda z)=z^Mg_{M}(w),     \label{eq:7}
\end{equation}
with 
\begin{equation}
	g_{M}(w)=\sum_{m=0}^{M}w^m c_{m}     \label{eq:8}
\end{equation}
and 
\begin{equation}
	c_{m}=\lambda^{M-m}\,P_{M-m}.    \label{eq:9}
\end{equation}
By choosing now the maximum of the ratio between two consecutive $c_{m}$
coefficients:
$$
\lambda = r[n_{\max}]=\max[P_{n}/P_{n+1}]=P_{n_{\max}}/P_{n_{\max}+1}
$$
one finds that the coefficients $c_{m}$ satisfy once again the EK theorem:
they are indeed real and positive and ordered
following the rule $c_{0}\geq \dots \geq c_{m}\geq \dots \geq c_{M}$. 
Therefore $g_{M}(w)$ differs from zero for $|w| < 1$ and the
zeroes of $g_{M}(w)$ lie necessarily in the region
$|w| \geq 1$.

Accordingly, all the zeroes of the truncated generating function
$G_{M}(z)$ fall in the annular region in the complex $z$-plane
centered  
at the origin and limited by $r[n_{\min}]$ and $r[n_{\max}]$. The
positive real axis is of course excluded.  

The mentioned theorem turns out to be quite useful when applied to the
class of power series distributions, 
defined as follows:
\begin{equation}
	P_{n}=c_{n} b^n P_{0}.     \label{eq:10}
\end{equation}
The corresponding $M$-truncated generating function is
\begin{equation}
	G_{M}(z)=A_{M}P_{0} \sum_{n=0}^{M}c_{n} b^nz^n=A_{M}P_{0} \sum_{n=0}^{M}c_{n} (bz)^n     \label{eq:11}
\end{equation}
where $A_{M}={(\sum_{n=0}^{M}P_{n})}^{-1}$ is a normalisation factor.
In the polynomial (\ref{eq:11}), the term $P_{0}$ can be neglected and
$bz$ replaced by the variable $u$, producing a new scaled  
generating function in $u$:
\begin{equation}
	H_{M}(u)=\sum_{n=0}^{M}c_{n} u^n     \label{eq:12}
\end{equation}
which ---as we will see in the following--- is of particular interest
for classifying different classes of events
described in terms of NB (Pascal) MD with different parameters.
In fact, applying the above considerations to the generating function
\begin{equation}
	G^{[NB]}(z)=\sum_{n=0}^{\infty}P_{n}^{[NB]}\,z^n=\left(\frac{k}{k+\bar{n}(1-z)}\right)^k     \label{eq:13}
\end{equation}
with 
\begin{equation}
	P_{n}^{[NB]}=\frac{k(k+1) \dots (k+n-1) {\bar{n}}^n k^k}{n! {(\bar{n}+k)}^{n+k}}  ,   \label{eq:14}
\end{equation}
leads to the $M$-truncated version of $G^{[NB]}(z)$:
\begin{equation}
	G_{M}^{[NB]}(z)=A_{M}^{[NB]}\sum_{n=0}^{M}P_{n}^{[NB]} z^n=A_{M}^{[NB]}\sum_{n=0}^{M}\prod_{i=1}^{n}\left(\frac{k+i-1}{i}\right)P_{0} b^n z^n  
	\label{eq:15}
\end{equation}
with $b =\frac{\bar{n}}{\bar{n}+k}$. 
Notice that in (\ref{eq:13}), $G^{[NB]}(z)$ has a singularity for $z=1/b$.

By defining now in (\ref{eq:15})
\begin{equation}
c_{n}^{[NB]}=\prod_{i=1}^{n}\left(\frac{k+i-1}{i}\right)     \label{eq:16}
\end{equation}
and
$$
bz=u,
$$
the new renormalised generating function, neglecting the inessential factor $P_{0}$, becomes 
\begin{equation}
	H_{M}^{[NB]}(u)=A_{M}^{[NB]}\sum_{n=0}^{M}\prod_{i=1}^{n}\left(1+\frac{k-1}{i}\right) u^n=A_{M}^{[NB]}\sum_{n=0}^{M}c_{n}^{[NB]} u^n.   \label{eq:17}
\end{equation}
Notice that the not-truncated version $H^{[NB]}(u)$ of (\ref{eq:17})
has a singularity at $u=1$. 
Relation (\ref{eq:17}) has been used in its not-truncated form in
Ref.~\cite{thermo-clan} where $c_{n}^{[NB]}$ has been identified with the
$n$-particle \emph{canonical partition function}; $H_{M}^{[NB]}(u)$ is
in this framework the \emph{truncated grand canonical partition
	function}. 
It follows that if parameter $k$ is less than one (as in the case of
the third class of events in pp collisions at 14 TeV c.m.\ energy
\cite{hard class}), 
then the coefficients $c_{n}^{[NB]}$ are all real and positive, and
decrease starting from $c_{0}^{[NB]}$.
The conditions for the applicability of the EK theorem are therefore
also here satisfied, and the zeroes of the scaled generating  
function (\ref{eq:17}) lie all outside the circle of unit radius $|u| = 1$.

This situation should be contrasted with the application of the EK
theorem when the parameter $k$ is greater than one (as in the case of
the soft and semi-hard classes of events in pp collisions \cite{rassegna}).
In this case in fact, the coefficients $c_{n}^{[NB]}$ are in
increasing order
and the zeroes lie all inside the circle of unit radius. 

As indicated in \cite{Marden} and shown in \cite{Brooks}, for the
(not-rescaled) NB (Pascal) MD Eq.~(\ref{eq:15}), one finds
\begin{equation}
	r[n]=\frac{P_{n}^{[NB]}}{P_{n+1}^{[NB]}}=\frac{1}{b}\left(\frac{n+1}{n+k}\right);     \label{eq:18}
\end{equation}
thus all zeroes of the algebraic equation of the
$M$-truncated generating function (\ref{eq:3}) lie inside the annular
region of the complex $z$-plane delimited by
$r[n_{\min}]$ and $r[n_{\max}]$, with
\begin{equation}
	r[n_{\max}]=\frac{1}{b}\left(\frac{M+1}{M+k}\right)\leq | z | \leq r[n_{\min}]=\frac{1}{kb}\equiv\frac{1}{a}\;\;\;\;(k<1)     \label{eq:19}
\end{equation}
and
\begin{equation}
	r[n_{\min}]=\frac{1}{kb}=\frac{1}{a}\leq | z | \leq r[n_{\max}]=\frac{1}{b}\left(\frac{M+1}{M+k}\right)\;\;\;(k>1)  .   \label{eq:20}
\end{equation} 
Notice that $r[n] \rightarrow \frac{1}{b}$ when $n \rightarrow
 \infty$, and $r[n] \rightarrow 1$ when 
 $\bar n \gg k$ (i.e. $b\rightarrow 1$).  
In the rescaled version (\ref{eq:17}), 
$r[n]=\frac{n+1}{n+k}$: it tends to 1 when $n \rightarrow \infty$ with
 constant $k$. 
It is remarkable that for $k=1$, $r[n]$ is also equal to one and all
zeroes lie on the circle of unit radius, the $u=1$ point being
excluded (see Figures 1 and 2). 

The role of parameter $a$ in the case of the NB (Pascal) MD is
underlined by the fact that the sum of the inverse of the zeroes of
the algebraic equation (\ref{eq:3}) is given by
\begin{equation}
	\sum_{i=1}^{M}\frac{1}{z_{i}}=-\frac{P_{1}}{P_{0}}.     \label{eq:21}
\end{equation}
Recall in fact that 
$$
\prod_{i=1}^{M}z_{i}={(-1)}^M\frac{P_{0}}{P_{M}}
$$
and also that the sum of ($n-1$) by ($n-1$) product of zeroes equals to 
$$
{(-1)}^{n-1}\frac{P_{M-(n-1)}}{P_{M}};
$$
for example for $M=3$ one gets
$$
z_{1}z_{2}+z_{1}z_{3}+z_{2}z_{3}={(-1)}^2 \frac{P_{1}}{P_{3}}.
$$
It follows that
$$
\sum_{i=1}^{M}\frac{1}{z_{i}}=\frac{{(-1)}^{M-1}P_{M-(M-1)}/P_{M}}{{(-1)}^MP_{0}/P_{M}}=-\frac{P_{1}}{P_{0}}.
$$
Notice that for the NB (Pascal) MD, Eq.~(\ref{eq:21}) becomes 
\begin{equation}
	\sum_{i=1}^{M}\frac{1}{z_{i}}=-b\,k=-\frac{\bar{n}k}{\bar{n}+k}=-a.     \label{eq:22}
\end{equation}
Equation (\ref{eq:21}) is indeed a \emph{general property} of
polynomials and is \emph{independent} of the value of $M$.
In the $u$-rescaled version, one has
\begin{equation}
	\sum_{i=1}^{M}\frac{1}{u_{i}}=-k.     \label{eq:23}
\end{equation} 
In addition, it should be pointed out that the NB parameters of the
coefficients of the $M$-truncated generating function (\ref{eq:15}),
namely $\bar n_{M}$ and $k_{M}$,  
can be calculated directly from the zeroes of equation (\ref{eq:3}). One finds:
\begin{equation}
	\bar n_{M}=\sum_{i=1}^{M}\frac{1}{1-z_{i}}     \label{eq:24}
\end{equation} 
and
\begin{equation}
	k_{M}=-\frac{{\left(\sum_{i=1}^{M}\frac{1}{1-z_{i}}\right)}^2}{\sum_{i=1}^{M}{\left(\frac{1}{1-z_{i}}\right)}^2}.     \label{eq:25}
\end{equation} 
In the limit $M \rightarrow \infty$, the parameters $\bar n_{M}$ and
$k_{M}$ coincide of course with the standard parameters of the
not-truncated distributions.  

One question still to be answered concerns the multiplicity of
zeroes of the algebraic equation (\ref{eq:3}), both in general and in the specific
case of the NB (Pascal) MD. 
Let us consider two polynomials $S(z)$ and $T(z)$ of degree $m$ and
$n$ respectively:  
$$
S(z)=s_{0}z^{m}+s_{1}z^{m-1}+ \dots +s_{m} \;\;\;\;\; (s_{0} \neq 0)
$$
and 
$$
T(z)=t_{0}z^{n}+t_{1}z^{n-1}+ \dots +t_{n} \;\;\;\;\; (t_{0} \neq 0)
$$ 
It is known that $S(z)$ and $T(z)$ possess a not constant common factor
iff the determinant 
$$
R_{S,T}=\det\left(
\begin{array}{ccccccccccc}
s_{0} & s_{1} & \cdots & \cdots & \cdots & s_{m-1} & s_{m} & 0 & \cdots & 0 & 0 \\ 
0 & s_{0} & s_{1} & \cdots & \cdots & \cdots & s_{m-1} & s_{m} & 0 & \cdots & 0 \\
\vdots & \vdots & \vdots & \vdots & \vdots & \vdots & \vdots & \vdots & \vdots & \vdots & \vdots \\
\vdots & \vdots & \vdots & \vdots & \vdots & \vdots & \vdots & \vdots & \vdots & \vdots & \vdots \\
0 & \cdots & \cdots & 0 & s_{0} & s_{1} & \cdots & \cdots & \cdots & s_{m-1} & s_{m} \\
t_{0} & t_{1} & \cdots & \cdots & t_{n-1} & t_{n} & 0 & \cdots & \cdots & \cdots & 0 \\
0 & t_{0} & t_{1} & \cdots & \cdots & t_{n-1} & t_{n} & 0 & \cdots & \cdots & 0 \\
0 & 0 & t_{0} & t_{1} & \cdots & \cdots & t_{n-1} & t_{n} & 0 & \cdots & 0 \\
\vdots & \vdots & \vdots & \vdots & \vdots & \vdots & \vdots & \vdots & \vdots & \vdots & \vdots \\
\vdots & \vdots & \vdots & \vdots & \vdots & \vdots & \vdots & \vdots & \vdots & \vdots & \vdots \\
0 & 0 & \cdots & \cdots & 0 & t_{0} & t_{1} & \cdots & \cdots & t_{n-1} & t_{n} \\
\end{array}
\right)
$$ 
is equal to zero. 

It follows that, called $S\,'$ the first derivative of the polynomial
$S(z)$ with respect to $z$, $S(z)$ will have at least one double 
root iff $R_{S,{S\,'}}=0$.
In order to be sure that the zeroes of a $M$-truncated generating
function $G_{M}(z)$ are all distinct, it is therefore sufficient to
show that none of the M roots $z_{i}$  is such that
$G_{M}(z_{i})={G\,'}_{M}(z_{i})=0$. 

By rewriting $P_{n}^{[NB]}(\bar n,k$) for simplicity in terms of
parameters $a$ and $b$ (see equation (\ref{eq:19}) for the definition
of $a$), one has  
\begin{equation}
	P_{n}^{[NB]}(a,b)=\frac{1}{n!}a(a+b) \dots (a+b(n-1))P_{0} ,    \label{eq:26}
\end{equation} 
and by properly normalising  $G_{M}(z)$ one finds, by induction,  
\begin{equation}
	R_{G_{M}^{[NB]},{G\,'}_{M}^{[NB]}}=A_{M}^{[NB]}\frac{a^M{[(a+b)(a+2b) \dots (a+Mb)]}^{M-1}{P_{0}}^{M-1}}{{(M!)}^{M-1}}.     \label{eq:27}
\end{equation} 

Since clearly $R_{G_{M}^{[NB]},{G\,'}_{M}^{[NB]}}$ is always positive for
$a$ and $b$ greater than zero, it follows that all the zeroes 
of the algebraic equation $G_{M}^{[NB]}(z)=0$ are distinct. 
Since all zeroes of $G_{M}^{[NB]}(z)=0$ are distinct, $M$ conditions are
needed to uniquely determine the set of zeroes.
For $k=1$ for instance, one should expect that
$$
\sum_{i=1}^{M}u_{i}=\sum_{i=1}^{M}{u_{i}}^2=\cdots=\sum_{i=1}^{M}{u_{i}}^M=-1.
$$
In fact, from (\ref{eq:16})
$$
\sum_{i=1}^{M}u_{i}=-\frac{c_{M-1}^{[NB]}}{c_{M}^{[NB]}}=-\frac{M}{M+k-1} \stackrel{k \rightarrow 1}{\longrightarrow} -1.
$$
The same limit is obtained for
$$
\sum_{i=1}^{M}{u_{i}}^2=\left(-\frac{c_{M-1}^{[NB]}}{c_{M}^{[NB]}}\right)^2-2\frac{c_{M-2}^{[NB]}}{c_{M}^{[NB]}} \stackrel{k \rightarrow 1}{\longrightarrow} -1
$$
as well as for the other powers up to $M$.
Notice that the same conditions are satisfied for $M \rightarrow
\infty$ and constant $k$:  consequently, also in this case the
zeroes stay on the circle of unit radius, as will be discussed in the
following.

\section{Maps of zeroes for different classes of events in the rescaled complex z-plane.} 

We propose in this Section to apply the previously discussed theorems
and remarks to the study, in the rescaled complex $z$-plane 
(i.e., in the complex $u$-plane), of maps of zeroes of the algebraic
equations obtained by truncating at integer value $M$ the NB (Pascal)
MD generating function. This distribution is
used in a statistical model of multiparticle production to describe
different classes of events (or substructures) in different
collisions. 
These substructures are for instance 2- and 3-jets events in
$e^{+}e^{-}$ annihilation or events without mini-jets (soft events)
and with mini-jets (semi-hard events) in pp collisions.
Recently \cite{hard class}, in pp collisions, the occurrence of a
third class of hard events at 14 TeV c.m.\ energy has been 
postulated also in the framework 
of the weighted superposition mechanism of different classes of
events, each described by a NB (Pascal) MD with characteristic NB
parameters. 
The occurrence of this third class of events was motivated 
by the fact that, by assuming both strong KNO scaling
violations and QCD behaviour, 
extrapolations of NB parameters in the LHC energy domain implied, in
the semi-hard sample of events, a surprising reduction in the average
number of clans \cite{rassegna}.

In order to deepen the analysis
of the properties of collective variables in terms of the maps of
complex zeroes and in order to study the possible occurrence
of a phase transition, we focus our attention on the three
mentioned classes of events at 14 TeV c.m.\ energy in pp collisions. 
Accordingly, the NB parameters used in the following are those
described in \cite{hard class}: $k_{soft}=7, \bar{n}_{soft}=40,
k_{semi-hard}=3.7,$ 
$\bar{n}_{semi-hard}=87, k_{hard}=0.1212, \bar{n}_{hard}=460$.

We find that for the soft and semi-hard components, the parameter $k$
is greater than one and the zeroes stay \emph{inside} the circle of unit radius
in the complex $u$-plane (see Figures 1a and 1b). For the hard
component, the parameter $k$ is smaller than one and all the
zeroes lie \emph{outside} 
the circle of unit radius $| u | =1$ (see Figure 1c).
In order to make our statements more transparent, in Figures 2a, 2b,
2c the behaviour of the three classes of events in the region around
the point $u=1$ has been magnified for different cut-offs but fixed
value of $k$. In Figure 3 the above maps are shown together close to
the point $u=1$,  with the same fixed cut-off for three values of
$k$ corresponding to the three classes of events.

Notice that for $k=1$ all the zeroes stay on the circle of unit radius
with the $u=1$ point excluded. In addition, when $M \rightarrow \infty$
the distribution of the zeroes of
the algebraic equation obtained from the truncated and rescaled
generating function converges towards the $u=1$ point 
for all classes of events with both $k>1$ and $k<1$.

As mentioned in the Introduction, in order to test the occurrence of a
phase transition one has to find the behaviour of the zeroes of the
grand canonical partition function $\mathcal{Q}$ in the
thermodynamical limit, i.e., with the size of the system going to
infinity. The discussion is based on the approach contained in
\cite{thermo-clan}, 
where the generating function $G(z)$ has been related to $\mathcal{Q}$ by
\begin{equation}
	\mathcal{Q} = [G(0)]^{-1}.     \label{eq:28}
\end{equation}
Relation (\ref{eq:28}) implies that the zeroes of the $M$-truncated
grand canonical partition function $\mathcal{Q}_{M}$ can be obtained
from the zeroes of the $M$-truncated generating function $G_{M}$
simply by rescaling those zeroes by a factor $b$ (which was identified
in \cite{thermo-clan} ---it should be stressed again--- with the fugacity).

The behaviour of the zeroes will then depend on the volume and the
number of particles: when the MD is of NB type, as we are using now,
it is sufficient to know 
the behaviour of the NB parameters. At present however, a realistic
link between $k$ and $\bar n$ and the system size (the volume) is
under investigation, 
hence postponed to a forthcoming paper. Here we limit ourselves to
explore what happens when the thermodynamical limit corresponds to $M$
(and $\bar n$) going to infinity with $k$ fixed.
In the first place we will check whether the EK theorem can be used to
confine the zeroes to a narrowing region of the complex
plane. Recalling Eq.~(\ref{eq:18}), we notice that for $k=1$ one has
$r[n]=\frac{1}{b}$. Thus minimum and maximum radii coincide and all
the zeroes of $G_{M}(z)$ lie on a circle of 
radius $1/b$, independently of the chosen cut-off $M$. Numerical
studies show that the opening around the positive real axis is reduced
when $M \rightarrow \infty$; this suggests that $z=1/b$ is an
accumulation point of zeroes. 
It means of course that in this scenario for $k=1$, the point $u=1$ is an
accumulation point of zeroes of the grand canonical partition
function: since $u$ on the real 
axis coincides with the fugacity, only for $b \rightarrow 1$ will a
phase transition take place. 
For $k>1$ the zeroes of the grand canonical partition function obey
the following relation  
\begin{equation}
	r[n_{\min}]=\frac{1}{k} \leq | u | \leq r[n_{\max}]=\frac{M+1}{M+k}     \label{eq:29}
\end{equation}
while for $k<1$ the two limits are exchanged. Such a region does not
shrink. We have to rely again on numerical root-finding, which still
indicates the closing in 
of the zeroes to $u=1$ (see Figures 1 and Figures 2), and at the same
time shows that they are much closer together than the bound given by
relation (\ref{eq:29}) allows. This feature again points in the direction of
the presence of a possible phase transition at $b \rightarrow 1$. 

In conclusion, the use of the rescaled complex variable $u=bz$ marks
visibly the different behaviour of the map of zeroes of the
hard class of events with respect to the maps of the
soft and semi-hard ones in our statistical model. In addition, it leads
to identify the coefficients $c_{n}^{[NB]}$  of Eq.~(\ref{eq:16}) with the
canonical partition function for a system of $n$ particles, and
$H_{M}(u)$ with the truncated grand canonical partition function of
the same system (in this framework $c_{1}=k$, i.e., the parameter $k$
coincides with the canonical partition function for a system with one
particle, see also \cite{thermo-clan}). It should also be  pointed out
that in  the rescaled version of this approach the NB parameter $k$ alone
controls the geometry of the distribution of the zeroes.
This is indeed a quite
general property  which is expected to be valid also in other
collisions and especially in those collisions involving events with
large numbers of particles, like heavy ion collisions.
Finally, it has been shown that the NB parameter $k$ is equal to minus
the sum of the inverse of the zeroes,
independently of the 
value of the cut-off $M$.

All these facts suggest to investigate further the thermodynamical
content of the NB parameter $k$, and especially its dependence on
temperature and volume, in view of a possible phase transition in the
$b \rightarrow 1$ limit. 
Being indeed the fugacity $b=e^{\mu/k_{B}T}$ (with $\mu$ the
chemical potential), one finds that $\mu$ goes to 0 in the
limit $b \rightarrow 1$: this is a
situation which is reached in a high temperature and high
density real boson gas and which is expected to occur in a quark-gluon
plasma.

\setlength{\parindent}{0pt}

  \begin{center}
  \mbox{\includegraphics[scale=0.9,bb=174 470 433 690]{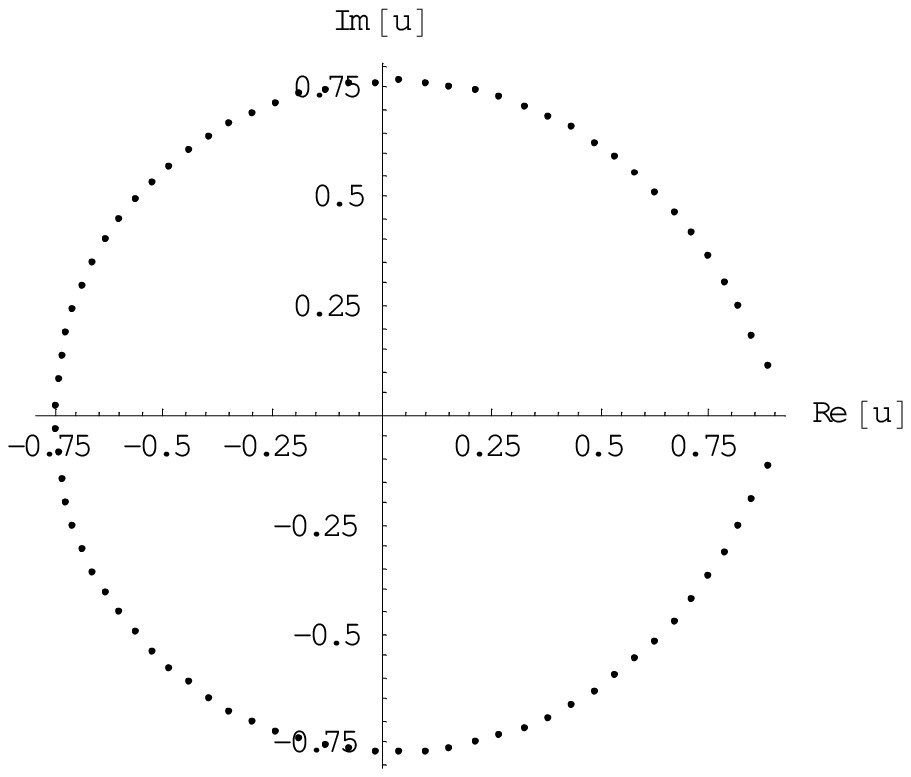}}
  \end{center}
Fig.~1a: rescaled map of zeroes, in the complex $u$-plane, of the NB
(Pascal) MD at 14 TeV c.m.\ energy with $k_{soft}=7$ and cut-off $M=80$
($M$ has been taken equal to $2\bar{n}_{soft}$). Zeroes \emph{lie
	inside} the circle of unit radius $| u | =1$. \\

  \begin{center}
  \mbox{\includegraphics[scale=0.9,bb=174 457 433 694]{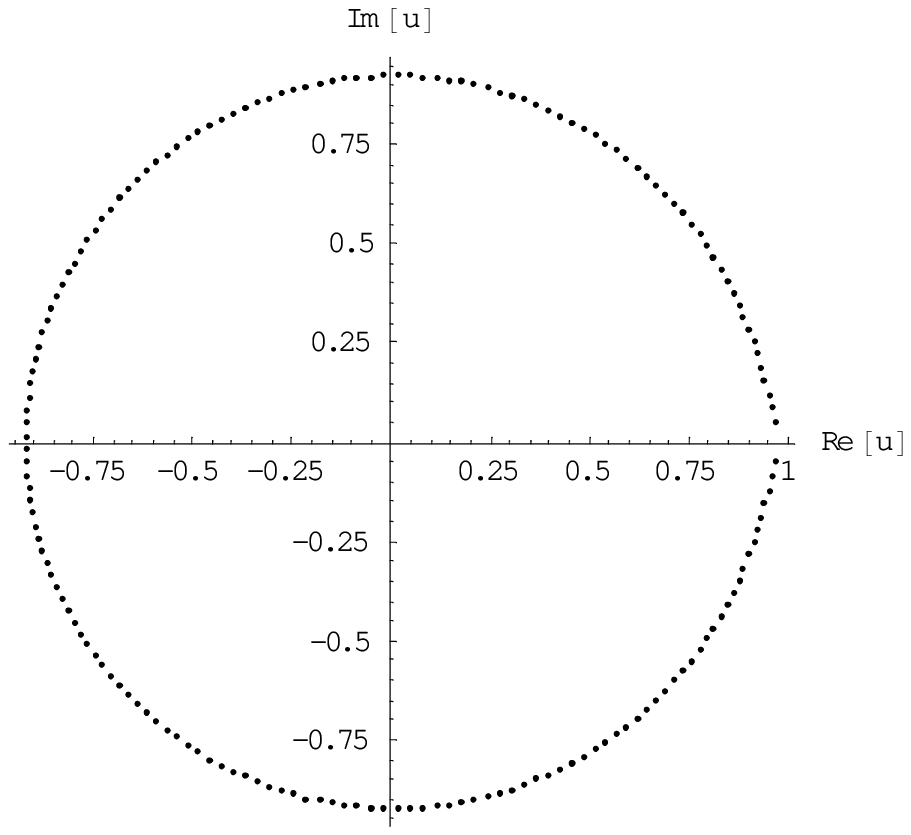}}
  \end{center}
Fig.~1b: rescaled map of zeroes, in the complex $u$-plane, of the  NB
(Pascal) MD at 14 TeV c.m.\ energy with $k_{semi=hard}=3.7$ and cut-off $M=174$
($M$ has been taken equal to $2\bar{n}_{semi=hard}$). Zeroes \emph{lie
	inside} the circle of unit radius $| u | =1$. \\

  \begin{center}
  \mbox{\includegraphics[scale=0.9,bb=180 462 427 694]{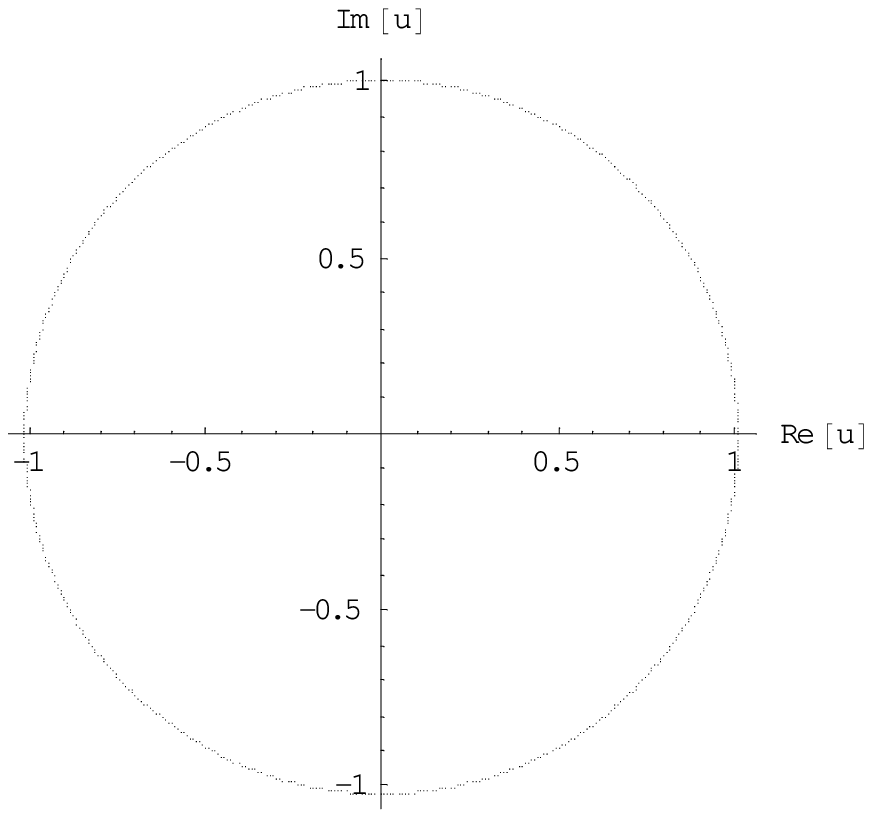}}
  \end{center}
Fig.~1c: rescaled map of zeroes, in the complex $u$-plane, of NB (Pascal)
MD at 14 TeV c.m.\ energy with $k_{hard}=0.1212$ and cut-off $M=600$.
Zeroes \emph{lie outside} the circle of unit radius $| u | =1$. \\

  \begin{center}
  \mbox{\includegraphics[scale=0.8,bb=111 113 496 695]{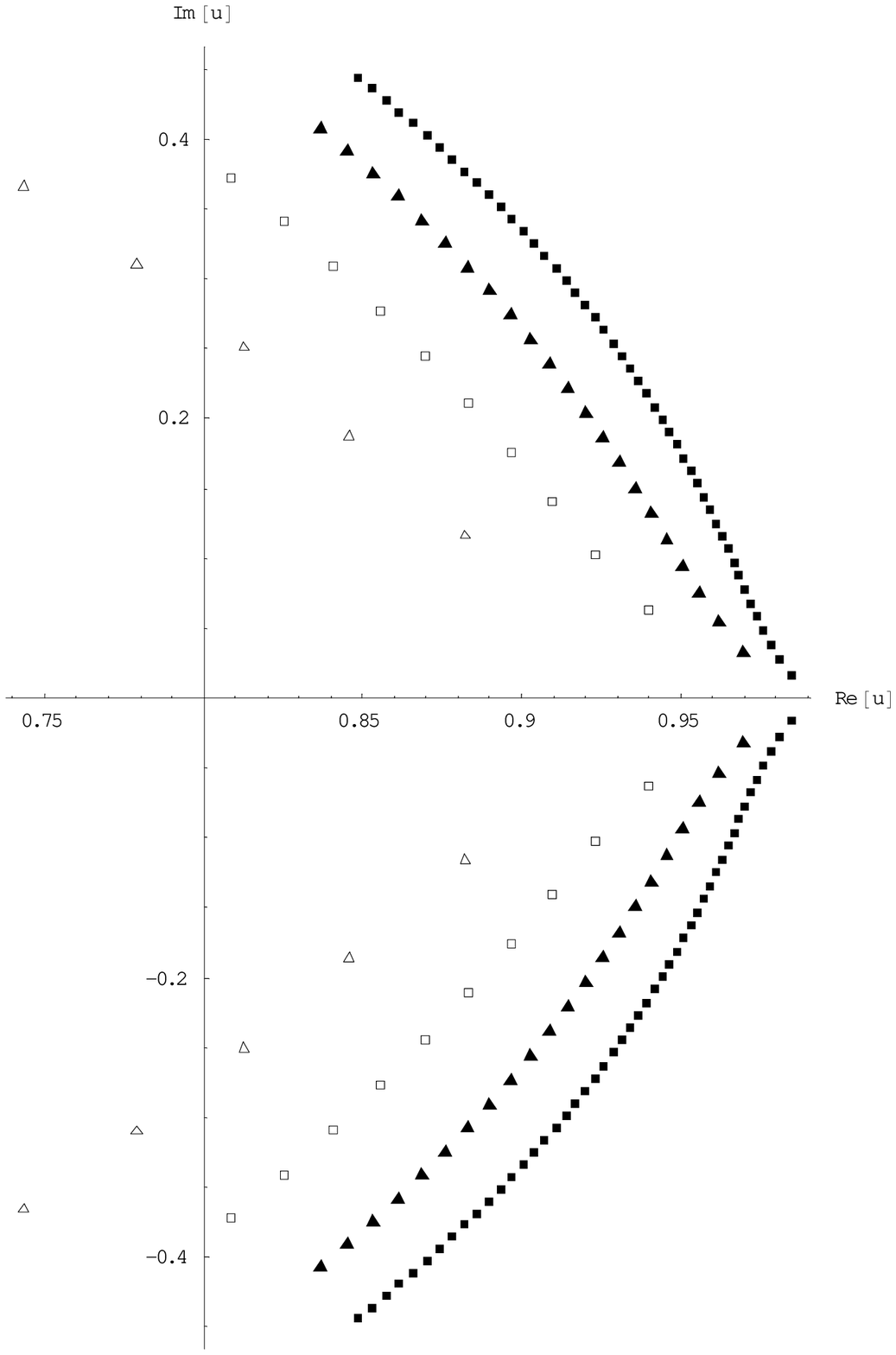}}
  \end{center}
Fig.~2a: the rescaled map of zeroes of the NB (Pascal) MD of Fig.~1a has
been magnified around the point $u=1$ for different cut-offs $M$ 
(white and black triangles correspond to $M=80$ and $M=320$
respectively, and white and black squares to $M=160$ and $M=640$). 
As $M$ increases, the point $u=1$ is shown to become an accumulation
point of zeroes \emph{from the inside}. \\

  \begin{center}
  \mbox{\includegraphics[scale=0.8,bb=111 125 496 709]{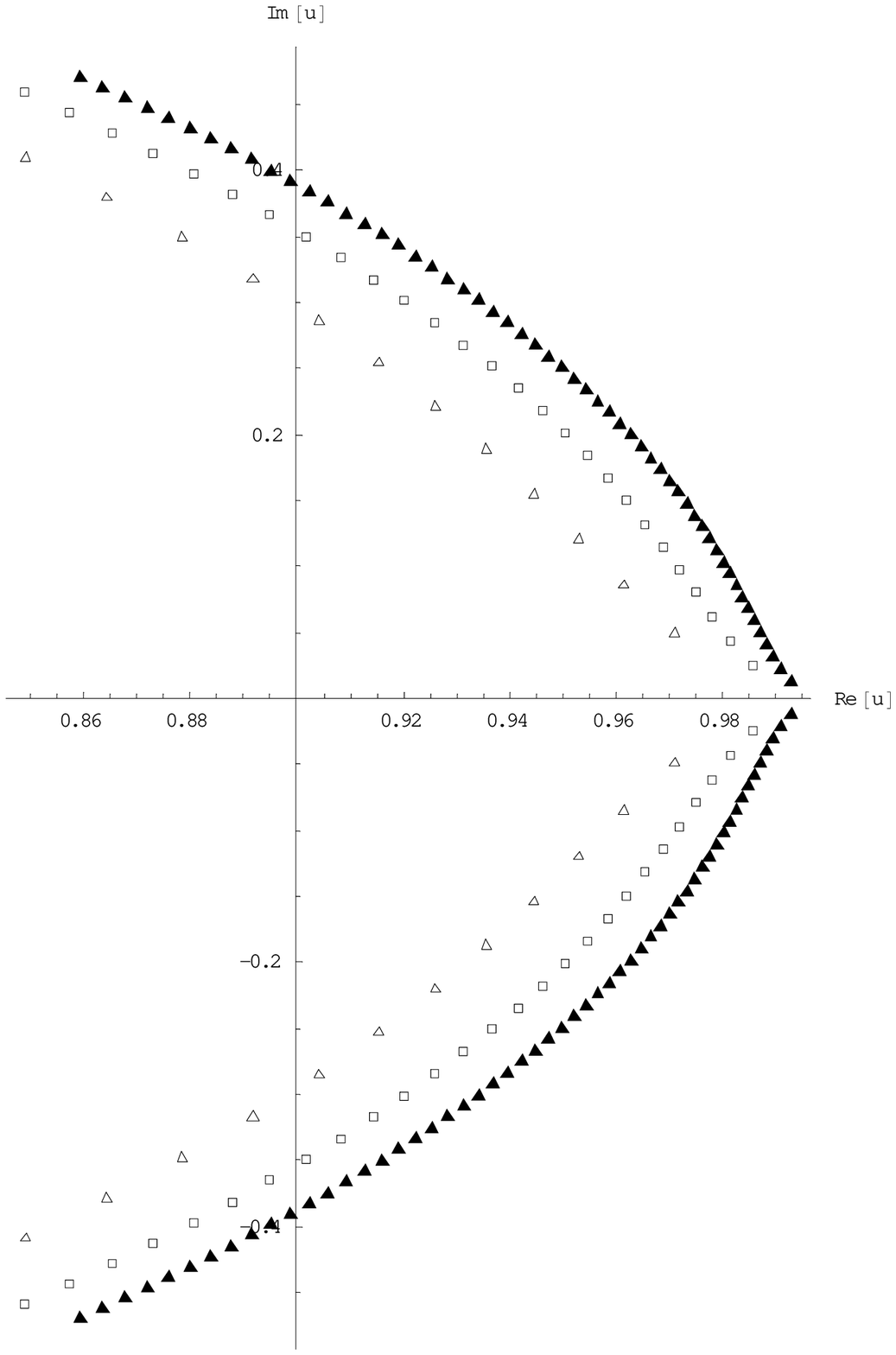}}
  \end{center}
Fig.~2b: the rescaled map of zeroes of the NB (Pascal) MD of Fig.~1b has
been magnified around the point $u=1$ for different cut-offs $M$ 
(white triangles correspond to $M=174$, white squares to $M=348$ and
black triangles to $M=696$). 
As $M$ increases, the point $u=1$ is shown to become an accumulation
point of zeroes \emph{from the inside}. \\

  \begin{center}
  \mbox{\includegraphics[scale=0.8,bb=111 125 496 709]{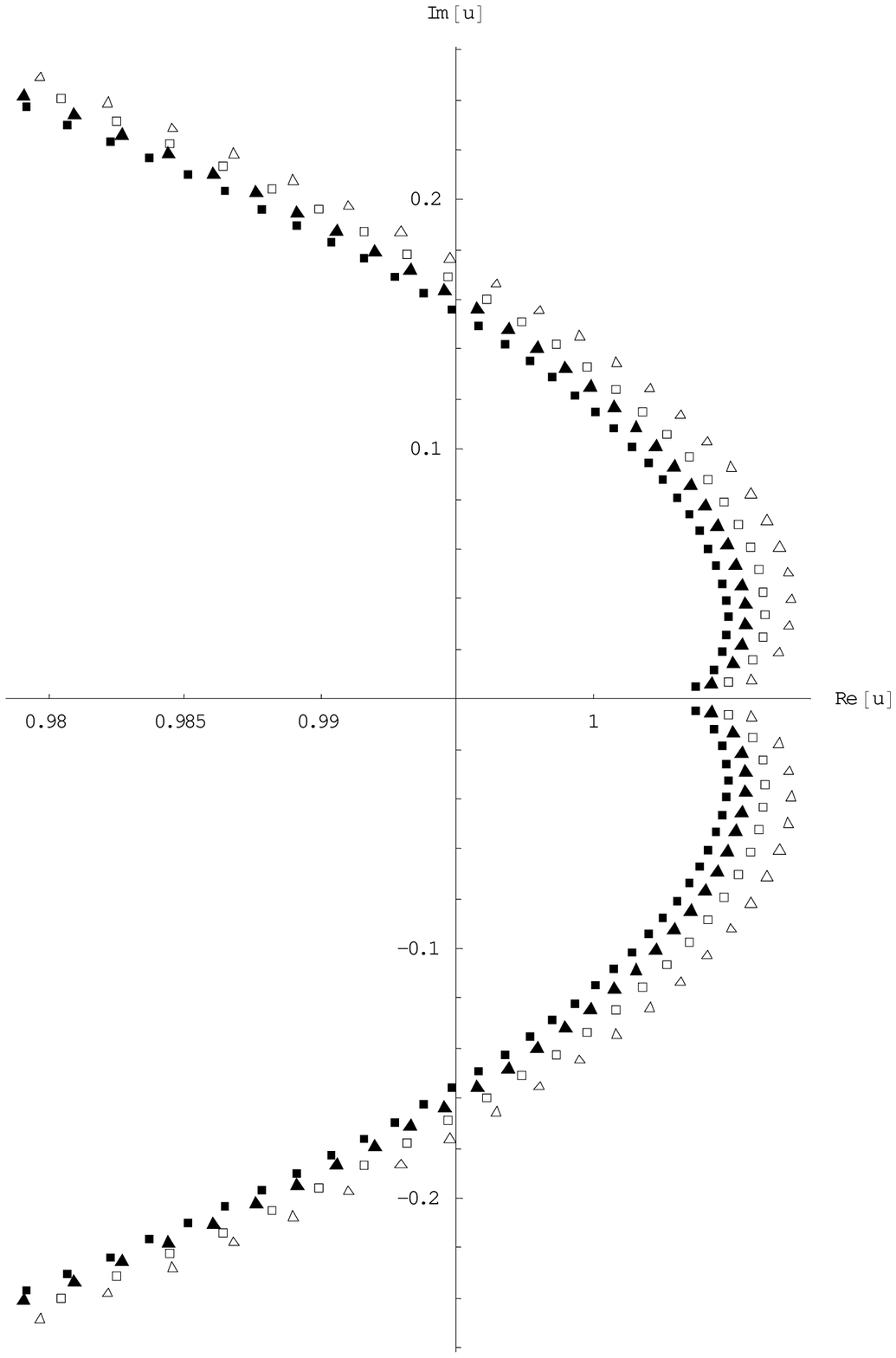}}
  \end{center}
Fig.~2c: the rescaled map of zeroes of the NB (Pascal) MD of Fig.~1c has
been magnified around the point $u=1$ for different cut-offs $M$ 
(white triangles and squares correspond to $M=600$ and $M=700$
respectively, and black triangles and squares to $M=800$ and
$M=920$). 
As $M$ increases, the point $u=1$ is shown to become an accumulation
point of zeroes \emph{from the outside}. \\

  \begin{center}
  \mbox{\includegraphics[scale=0.8,bb=135 188 472 696]{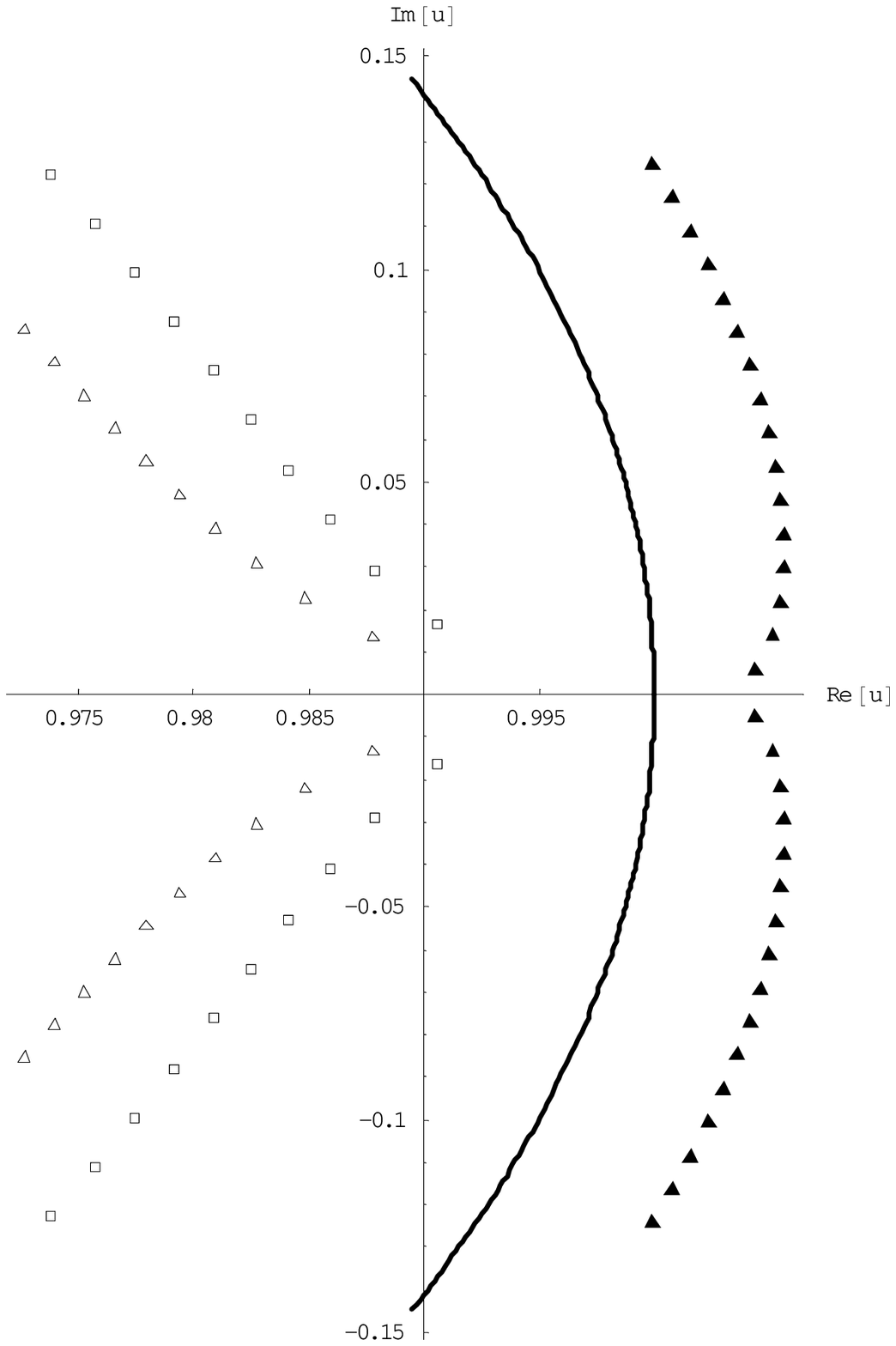}}
  \end{center}
Fig.~3: maps of zeroes of the NB (Pascal) MD are shown for different
values of the parameter $k$ and the same fixed cut-off $M=800$. 
White triangles correspond to $k_{soft}$ of Fig.~1a, white squares to
$k_{semi-hard}$ of Fig.~1b and black triangles to $k_{hard}$ of Fig.\
1c.  
The full line describes the circle of unit radius $| u | =1$.
The figure illustrates that  ---as discussed in the text---
the point $u=1$ is an accumulation point of zeroes \emph{from the inside} (for
$k>1$) and \emph{from the outside} (for $k<1$.)


\begin{thebibliography}{7}
   \bibitem{thermo-clan} A.Giovannini, S.Lupia and R.Ugoccioni, Phys.Rev.D 65, 094028 (2002)
   \bibitem{Lee-Yang}    C.N.Yang and T.D.Lee, Phys.Rev. 87, 404
   (1952); 87, 410 (1952)
	 \bibitem{Biebl} K.J.\ Biebl and J.\ Wolf, Nucl.Phys. B 44, (1972)
	 301
	 \bibitem{Discrete} B.\ Andersson, G.\ Gustafson and J.Samuelsson, 
		 Nucl.Phys .B 463, (1996) 217
   \bibitem{De Wolf}     E. De Wolf, in Proceedings of the XIV Int. Symp. on Multiparticle Dynamics, Eds. A.Giovannini, S.Lupia, R.Ugoccioni,
                         World Scientific, Singapore 1995, p.15
   \bibitem{Brooks}      T.C.Brooks, K.L.Kowalski and C.C.Taylor, Phys.Rev.D 56:5857--5861, 1997
   \bibitem{Marden}      M.Marden, \emph{Geometry of polynomials} (Mathematical Surveys Number 3, American Mathematical Society,
                         Providence, Rhode Island, 1966, p.136--137)
   \bibitem{rassegna}    A.Giovannini and R.Ugoccioni, Phys.Rev.D 59, 094020 (1999) and Phys.Rev.D 60, 074077 (1999)
   \bibitem{hard class}  A.Giovannini and R.Ugoccioni, Phys.Rev.D 68, 034009 (2003)
\end{thebibliography}
\end{document}